\documentclass[preprintnumbers,amsmath,amssymb,aps,prl,floatfix,lengthcheck,superscriptaddress]{revtex4}
\usepackage{graphicx}
\usepackage{braket}
\usepackage{upgreek}
\usepackage{color}

\begin{document}

\title{Observation of Rydberg Blockade Induced by a Single Ion}

\author{F. Engel}
\author{T. Dieterle}
\author{T. Schmid}
\author{C. Tomschitz}
\author{C. Veit}
\author{N. Zuber}
\author{R. L\"{o}w}
\author{T. Pfau}
\author{F. Meinert}
\affiliation{5. Physikalisches Institut and Center for Integrated Quantum Science and Technology, Universit\"{a}t Stuttgart, Pfaffenwaldring 57, 70569 Stuttgart, Germany}
\date{\today}

\begin{abstract}
We study the long-range interaction of a single ion with a highly excited ultracold Rydberg atom and report on the direct observation of ion-induced Rydberg excitation blockade mediated over tens of micrometer distances. Our hybrid ion-atom system is directly produced from an ultracold atomic ensemble via near-threshold photo-ionization of a single Rydberg excitation, employing a two-photon scheme which is specifically suited for generating a very low-energy ion. The ion's motion is precisely controlled by small electric fields, which allows us to analyze the blockade mechanism for a range of principal quantum numbers. Finally, we explore the capability of the ion as a high-sensitivity single-atom-based electric field sensor. The observed ion - Rydberg-atom interaction is of current interest for entanglement generation or studies of ultracold chemistry in hybrid ion-atom systems.
\end{abstract}

\maketitle

Ultracold Rydberg atoms have recently been shown to provide a highly flexible platform for quantum simulation of long-range interacting many-body systems \cite{Schauss2012,Labuhn2016,Bernien2017}, generation of nonclassical photonic states \cite{Peyronel2012,Firstenberg2013}, or quantum information processing applications \cite{Isenhower2010,Jau2016,Tiarks2016}. A central aspect in this context is the Rydberg blockade phenomenon, which inhibits the simultaneous excitation of two close-by atoms into Rydberg states as a consequence of strong Rydberg-Rydberg interaction \cite{Lukin2001}. A similar concept applies to hybrid systems of Rydberg atoms and ions, for which strong mutual interaction may lead to single charge-induced blockade phenomena mediated over macroscopic distances \cite{Secker2016}. Collisions of highly excited atoms with ions have been studied early on with atom-beam experiments \cite{Gallagher1994}. More recently, their interaction affects quantum optics applications based on room temperature vapors \cite{Weller2016}.

In the context of trapped ions and ultracold atoms \cite{Grier2009,Zipkes2010,Schmid2010,Tomza2017}, where motional degrees of freedom are exquisitely controlled at the single particle level, strong coupling of ions to Rydberg atoms has been proposed for generating ion-atom entanglement \cite{Secker2016} and for controlling cold collisions, chemistry, or charge mobilities in ion-atom mixtures \cite{Secker2017,Cote2000}. However, the observation of an ion-induced Rydberg blockade so far remained elusive. The major obstacle to probe ion - Rydberg-atom interaction in traditional hybrid settings, based on radio-frequency ion traps and Rydberg states excited from trapped neutral ensembles, are trap-induced lineshifts on the Rydberg state, which are challenging to distinguish from interaction effects. Moreover, systems of trapped Rydberg ions are essentially unamenable to ion - Rydberg-atom interaction due to the cancellation of Coulomb and trapping fields at the ions' equilibrium positions \cite{Feldker2015,Higgins2017}.

In this Letter, we demonstrate excitation blockade of a single Rydberg atom by a single low-energy ion. The ion is produced from a single Rydberg excitation in an ultracold sample via a novel optical two-photon ionization scheme. Our approach provides precise spatial and motional control of the initially ultracold free ion, which constitutes the key ingredient to probe the blockade mechanism. Finally, we exploit the blockade to use the ion as a precise quantum probe for small electric stray fields.

\begin{figure}[!ht]
\centering
	\includegraphics[width=\columnwidth]{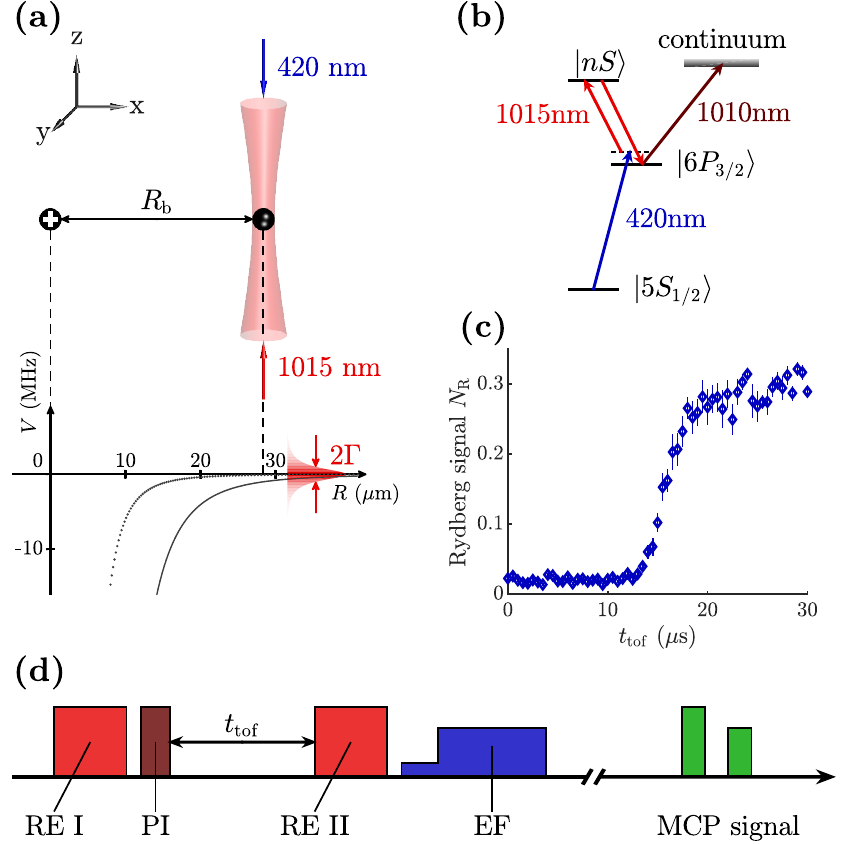}
	\caption{Concept for observing ion-induced Rydberg blockade. (a, top) Single Rydberg atom excitation is probed in the presence of an ion at distance $R$, determined by the ion's trajectory in an external electric field $E$ after time-of-flight $t_{\rm{tof}}$. Focused lasers provide spatial control of ion and Rydberg atom production. (a, bottom) Pair-interaction potential $V(R)$ between ion and Rydberg atom for $n=71$ (dotted line) and $n=100$ (solid line) as a function of $R$. The shaded red curve indicates the Rydberg excitation bandwidth $\Gamma$. (b) Rydberg excitation and photo-ionization scheme. (c) Mean number of detected Rydberg atoms $N_{\rm{R}}$ as a function of the ion's time-of-flight $t_{\rm{tof}}$ in the external stray electric field.  Error bars indicate one standard deviation obtained from averaging over 19 realizations. (d) Experimental sequence consisting of first Rydberg excitation (RE I), ion production via photo-ionization (PI), second Rydberg excitation (RE II), two-step electric field pulse (EF) for selective ion and Rydberg atom detection, and temporarily separated detection events on the MCP.}
	\label{Fig1}
\end{figure}

Consider a pair of a highly excited Rydberg atom and an ion in its electronic ground-state separated by a distance $R$. We focus on $nS_{1/2}$ Rydberg states of $^{87}\textrm{Rb}$ with principal quantum number $50 \lesssim n \lesssim 100$. At sufficiently large internuclear distance, where the Coulomb field of the ion has dropped well below the Inglis-Teller limit, the pair interaction is described by the long-range and isotropic polarization potential \cite{Secker2016,Hahn2000}
\begin{equation}
 V(R) = - C_4/(2 R^4) \, .
\label{eq1}
\end{equation}
The strength of interaction is determined by the Rydberg atom's polarizability $\alpha_{nS} = (4 \pi \epsilon_0 / e)^2 C_4$. The exaggerated polarizability of high-$n$ Rydberg states causes sizeable interaction energies of several MHz over tens of $\mu$m distances as depicted in Fig.~\ref{Fig1}(a) at the example of two different principal quantum numbers. Consequently, the close-by ion may affect the attempt to laser excite the atom from its ground state to the Rydberg state. More specifically, excitation blockade is expected to occur at internuclear distances for which the interaction shift exceeds the excitation bandwidth $\Gamma$. In close analogy to the pair interaction of two Rydberg atoms, this defines a blockade sphere surrounding the ion with a radius given by
\begin{equation}
R_{\rm{b}} = (C_4/(2 \Gamma))^{1/4} \, .
\label{eq2}
\end{equation}

In our experiment, we probe this blockade sphere starting from an ultracold ensemble of typically $1.2 \times 10^5$ $^{87}\textrm{Rb}$ atoms prepared in the $|5 S_{1/2}, F=2, m_F=2\rangle$ hyperfine state and held in a crossed-beam optical dipole trap at a temperature of about $1 \, \mu \rm{K}$. The trap frequencies are $(\omega_x,\omega_y,\omega_z)=2 \pi \times (160,110,710) \, \rm{Hz}$. From the trapped sample, a single atom is excited to the $|nS_{1/2}\rangle$ Rydberg level via a $500 \, \rm{ns}$ long two-photon excitation pulse incorporating the intermediate $6P_{3/2}$ level at a detuning of $+160$ MHz. The laser beams for Rydberg excitation (420 and 1015 nm wavelength) counter-propagate along the vertical $z$-direction and are circularly polarized to address the $m_J=+1/2$ Zeeman substate. The 1015 nm laser is focused into the sample through a high-NA aspheric lens to a beam waist of about $1.8 \, \mu \rm{m}$, providing spatial control over the position of the created Rydberg atom. Note that for the high-lying Rydberg states investigated in this work, Rydberg-Rydberg blockade allows us to suppress multiple excitations in the sample.

The Rydberg atom is then photo-ionized within 200 ns to produce a single low-energy ion. For this, we apply a two-photon V-type ionization scheme employing a first laser tuned in resonance with the $6P_{3/2}$ level and a second laser at about 1010.2 nm wavelength, which provides just enough energy to reach the ionization continuum (Fig.~\ref{Fig1}(b)) \cite{SM,Schmid2018}. The calculated excess kinetic energy of the ion is below $k_B \times 10 \, \mu \rm{K}$ or $0.86 \, \rm{neV}$. Consequently, the ion trajectory after production is entirely governed by residual electric stray fields, which in the course of the experiment are compensated to a level of $\sim 1.5 \, \rm{mV/cm}$ via Stark spectroscopy at $n=133$ \cite{Osterwalder1999}. The slow motion of the single ion in the electric field allows for a first probe of the ion-induced Rydberg blockade. Specifically, we let the ion propagate for a variable time-of-flight $t_{\rm{tof}}$, during which it increases its distance from the initial excitation region (depicted laser focus in Fig.~\ref{Fig1}(a)). Subsequently, a second Rydberg excitation pulse with the same pulse parameters as above is applied to probe the success of Rydberg atom creation in the presence of the nearby ion.

The ion and the Rydberg atom stemming from the second excitation pulse are detected individually on a microchannel plate detector (MCP). They are distinguished by their flight time to the MCP using a two-step extraction electric field pulse. A first field step of $1.5 \, \rm{V/cm}$ applied for $5 \, \mu \rm{s}$ only accelerates the ion towards the detector. In a second step the field is set to $115 \, \rm{V/cm}$ for $15 \, \mu \rm{s}$ to field-ionize and subsequently detect the Rydberg atom. The described pulse sequence, summarized in Fig.~\ref{Fig1}(d), is repeated 500 times in the same ensemble of atoms to gain statistics. The individual detection of the ion and Rydberg atom allows us to post-select events for which ion creation and detection was successful \cite{SM}. Fig.~\ref{Fig1}(c) shows the mean number of detected Rydberg atoms $N_{\rm{R}}$ for these events as a function of $t_{\rm{tof}}$ recorded for principal quantum number $n=90$. For short times $t_{\rm{tof}} \lesssim 15 \, \mu \rm{s}$, we observe an almost total suppression of excited Rydberg atoms. Only for longer time-of-flight, when the ion is far enough from the Rydberg excitation region, $N_{\rm{R}}$ reaches the value measured in the absence of the ion.

The observed suppression is attributed to the excitation blockade induced by the single ion. Indeed, the value of $t_{\rm{tof}}$ for which $N_{\rm{R}}$ rapidly increases is understood also quantitatively. For this, consider the theoretical value of the blockade radius $R_{\rm{b}} = 23.1 \, \mu \rm{m}$ for $n=90$ and $\Gamma = 1.09(1) \, \rm{MHz}$. Note that in all our experiments, the excitation bandwidth $\Gamma$ is determined from the independently measured spectral width, and is Fourier-limited by the finite duration of the Rydberg excitation pulse \cite{SM}. Assuming a plain accelerated motion of the ion in a constant electric field $E_{\rm{stray}}$, the time-of-flight at which $N_{\rm{R}}$ increases suggests $E_{\rm{stray}} \sim 1.7 \, \rm{mV/cm}$, which is compatible with the level of stray field compensation discussed above.

\begin{figure}[t]
\centering
	\includegraphics[width=\columnwidth]{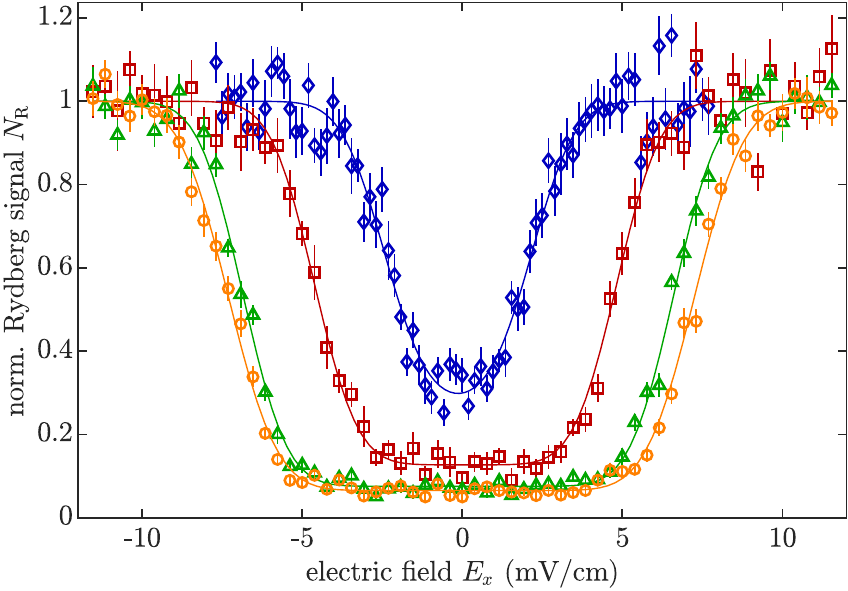}
	\caption{Ion-induced Rydberg blockade for different principal quantum numbers. Mean number of detected Rydberg atoms $N_{\rm{R}}$, normalized to the value obtained in the absence of the ion, as a function of the applied electric field $E_x$ for $n=51$ (diamonds), $n=71$ (squares), $n=90$ (triangles), and $n=100$ (circles). The ion's time-of-flight in the presence of $E_x$ is set to $t_{\rm{tof}} = 7 \, \mu \rm{s}$ for all data sets. Solid lines are error-function fits to the data to extract the blockade radius (see text for details). Error bars indicate one standard deviation obtained from averaging over typically 20 realizations.}
	\label{Fig2}
\end{figure}

Next, we provide a rigorous experimental determination of the blockade radius for a range of principal quantum numbers. Evidently, the motion of the ion in the small yet unknown stray electric field impedes a direct measurement of $R_{\rm{b}}$. Therefore, we slightly adapt the aforementioned experimental procedure and now apply a variable electric field $E_x$ along the $x$-direction throughout the entire sequence to control the ion trajectory during $t_{\rm{tof}}$. Here, the ion's time-of-flight is kept at a fixed value $t_{\rm{tof}} = 7 \, \mu \rm{s}$. As before, we investigate the post-selected Rydberg signal $N_{\rm{R}}$ measured in the presence of a detected ion. The obtained results as a function of $E_x$ are shown in Fig.~\ref{Fig2} for four values of $n$. For sufficiently large $E_x$, we find that $N_{\rm{R}}$ settles to the value obtained in the absence of the ion. This indicates that the distance $R$ between the ion and the Rydberg excitation region has increased beyond $R_{\rm{b}}$ during $t_{\rm{tof}}$ due to the applied electric field. For small values of $E_x$, strong Rydberg excitation blockade is observed as $R$ is still well below $R_{\rm{b}}$ after $t_{\rm{tof}}$. The reduced contrast for the data set at $n=51$ is most likely due to the finite Rydberg excitation and photo-ionization volume determined by the size of the focused laser beams. The transition between the blockaded and the non-blockaded situation is marked by a pronounced and comparatively sharp $n$-dependent edge in the signal, which allows us to extract $R_{\rm{b}}$ from the data.

\begin{figure}[!t]
\centering
	\includegraphics[width=\columnwidth]{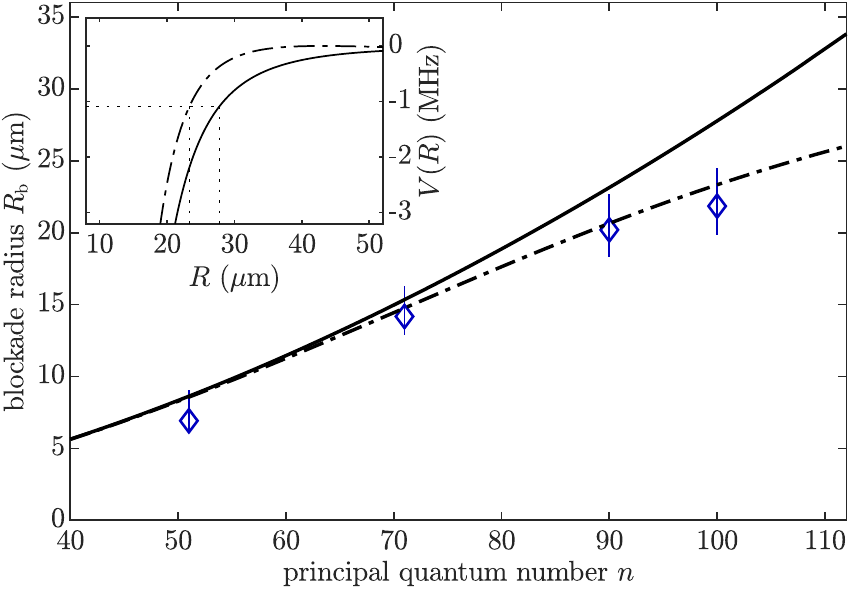}
	\caption{Blockade radius for different principal quantum numbers. Symbols show the measured blockade radius $R_{\rm{b}}$ extracted from the data in Fig.~\ref{Fig2} as a function of $n$. Error bars include the effects of residual stray fields and finite pulse length for photo-ionization and Rydberg excitation. The solid line shows the prediction according to Eq.~\ref{eq2}. The dash-dotted line is the result of an extended theoretical description taking into account the effect of the applied electric field $E_x$ during the measurement. The inset shows the bare (solid line) and the modified (dash-dotted line) $V(R)$ in the presence of the field $E^{\star}$ for $n=100$. Dotted lines indicate the shift of $R_{\rm{b}}$ for the experimental value of $\Gamma = 1.09(1) \, \rm{MHz}$.}
	\label{Fig3}
\end{figure}

In order to determine $R_{\rm{b}}$, we fit each data set with an error function of the form $\propto {\rm{erf}}((E_x - E^{\star})/w)$, symmetrized to evaluate the experimental signal at positive and negative field values. The fitting procedure delivers the center position of the observed edge $E^{\star}$ for each value of $n$. Given the accelerated motion of the ion in the applied field, the blockade radius is directly associated with $E^{\star}$ via $R_{\rm{b}} = e E^{\star}  t_{\rm{tof}}^2 / (2m)$, where $m$ denotes the mass of the $^{87}$Rb atom. Note that we have tested this simple relation against numerical simulations of the single Rydberg atom excitation dynamics in the presence of the moving ion and found excellent agreement \cite{SM}. The experimentally obtained values for the blockade radius as a function of $n$ are depicted in Fig.~\ref{Fig3}. Error bars account for two dominant sources of experimental uncertainty. For low principal quantum numbers (specifically for $n=51$), the comparatively small value of $E^{\star}$ starts to compete with the residual stray fields. This effect quickly diminishes with increasing $n$. The error is then dominated by the finite pulse length for photo-ionization and Rydberg excitation, thereby causing a statistical uncertainty on $t_{\rm{tof}}$.

The obtained data for $R_{\rm{b}}$ are compared to the prediction of Eq.~\ref{eq2} (solid line in Fig.~\ref{Fig3}). While the expected increase of the blockade radius with the principal quantum number is clearly observed, a systematic shift to smaller values of $R_{\rm{b}}$ is identified with increasing $n$. The deviation is attributed to the presence of the electric field $E_x$ that controls the motion of the ion. The maximum values for $E_x$ have been carefully chosen to ensure a negligible Stark shift on the Rydberg line compared to our excitation bandwidth $\Gamma$ in the field-free case when no ion is present. However, the situation changes in the presence of the ion. In that case, $E_x$ has a significant effect on the blockade radius, which is most pronounced for large values of $n$.

To quantify this point, let us recall the origin of the pair-interaction $V(R)$, i.e. the energy shift of the Rydberg state in the Coulomb field of the ion. The field of the ion at $R_{\rm{b}}$ for $n=100$ is about $18 \, \rm{mV/cm}$ and thus on the order of $E_x$. Furthermore, the two field vectors always point in opposite direction at the position of the excited Rydberg atom, independent of the sign of $E_x$. Consequently, the applied external field significantly reduces the total field strength acting on the Rydberg state and thus leads to a systematic shift of $R_{\rm{b}}$ to smaller values (see inset to Fig.~\ref{Fig3}). The effect can be easily accounted for in a self-consistent way, which provides the corrected value of $R_{\rm{b}}$ and the corresponding $E^{\star}$ after a few iterations \cite{SM}. The obtained result is compared to the data  in Fig.~\ref{Fig3} (dash-dotted line) and agrees very well within the experimental error bars.

\begin{figure}[!t]
\centering
	\includegraphics[width=\columnwidth]{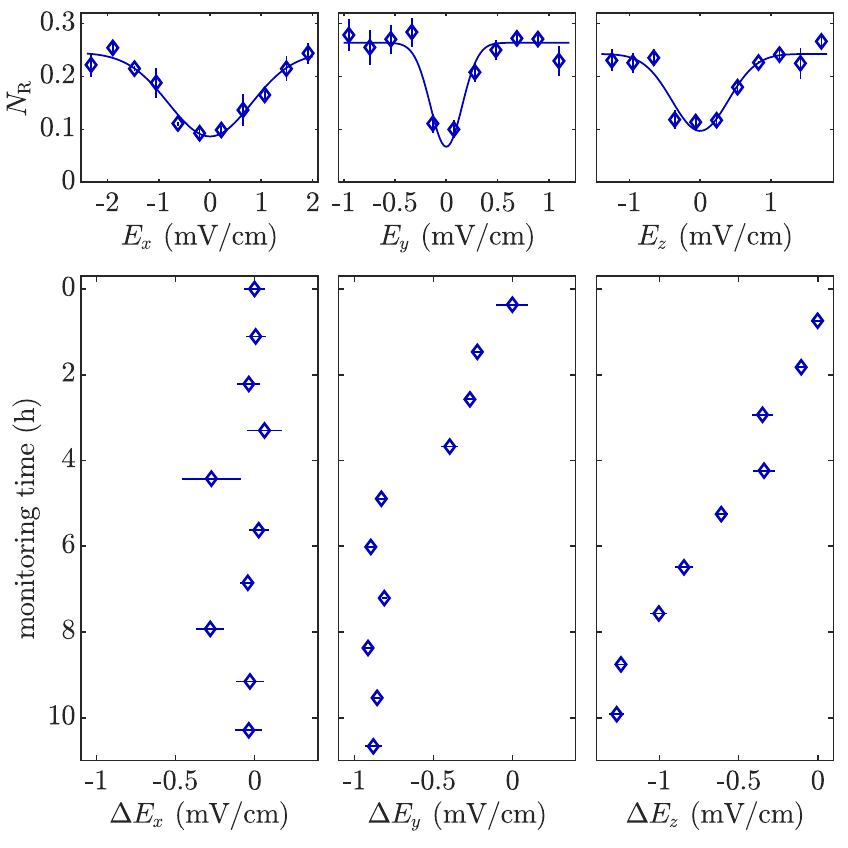}
	\caption{A single ion as an electric field probe. (top row) Mean number of detected Rydberg atoms $N_{\rm{R}}$ as a function of applied electric field $E_{i}$ in all spatial directions $i=\{x, y,z\}$ for $n=100$. Here, the ion's time-of-flight $t_{\rm{tof}} = 34 \, \mu \rm{s}$. Solid lines are Gaussian fits to the data to extract the center of the ion-induced blockade feature, which is identified with zero field strength for the respective spatial component. Error bars show one standard deviation obtained from averaging over typically four realizations. (bottom row) Temporal drift $\Delta E_{i}$ of the three electric field components. Symbols denote the change of the fitted centers obtained from scans as shown in the top row. Error bars indicate the $1\sigma$ confidence bound for the fit result.}
	\label{Fig4}
\end{figure}

Finally, we explore the potential of the ion-induced Rydberg blockade measurement as a sensitive tool for probing small electric fields. When increasing $t_{\rm{tof}}$ in the above measurement sequence, smaller field strengths are required to drag the ion out of the blockade volume. Consequently, a higher sensitivity to small electric fields can be achieved as evident in Fig.~\ref{Fig4} (top row), where we apply our protocol as before but now with $t_{\rm{tof}}=34 \, \mu \rm{s}$. The Rydberg signal $N_{\rm{R}}$ as a function of the applied field $E_{i}$ in all three spatial directions $i=\{x, y,z\}$ reveals the blockade feature around zero field strength with a strongly reduced width well below $2 \, \rm{mV/cm}$. The difference in the widths observed in the three directions is attributed to residual field gradients on the order of $1 \, \rm{mV/cm}$ across $10 \, \mu \rm{m}$, which start to affect the ion's trajectory for the very small homogeneous field components probed in this measurement. Extracting the center positions of the blockade dip allows for precise monitoring and compensation of temporal stray electric field changes at the level of $\lesssim 100 \, \mu \rm{V/cm}$. We have measured such field drifts over the course of several hours and show the results in Fig.~\ref{Fig4} (bottom row). Note that for the achieved level of control, we expect that even the detection of the ion's tiny excess energy from the photo-ionization process should be within reach when field gradients are carefully compensated.

Our measurements manifest that the produced ion can serve as a precise single-atom electric field sensor, which provides significant improvement for stray field compensation in our apparatus as compared to Rydberg Stark spectroscopy (\textit{cf.} the level of stray field compensation given above for $n=133$). The method should yield even higher sensitivity when further increasing $t_{\rm{tof}}$. In our experiment, however, we are currently limited by slow field drifts during data acquisition, which for a single data set in the top row of Fig.~\ref{Fig4} takes about 20 minutes.

In conclusion, we have studied a so far unexplored type of strong and long-range interaction in the realm of ultracold gases between an ion and a highly excited Rydberg atom. We have observed and quantified efficient excitation blockade for the Rydberg atom in the presence of the ion on the single-particle level by exploiting a novel photo-ionization scheme, suited for the production of extremely low-energy ions. The reported techniques may be readily applied in a dense gas or even a Bose-Einstein condensate for investigating and controlling ion-atom collisions and chemistry without the need of an ion trap. Combining the observed blockade with Rydberg-EIT might even allow for in-situ imaging of the ion kinetics \cite{Guenter2012,Guenter2013}. Selective photo-association of an ultralong-range Rydberg molecule \cite{Greene2000,Bendkowsky2009} prior to the photo-ionization pulse could provide ideal initial conditions for collision studies, with promising prospects for reaching the ion-atom quantum scattering regime \cite{Schmid2018}. Moreover, using the ion as a precise field sensor sets the stage for optical trapping of the single ion \cite{Huber2014,Lambrecht2017}, here starting from an ultracold Rydberg excitation.

We thank S. Weber, K. Kleinbach, R. Gerritsma, M. Dei{\ss}, and J. Hecker Denschlag for stimulating discussions.
We acknowledge support from Deutsche Forschungsgemeinschaft [Projects No. PF 381/13-1 and No. PF 381/17-1, the latter being part of the SPP 1929 (GiRyd)].
F. M. and C. V. acknowledge support from the Carl-Zeiss foundation. F. M. is indebted to the Baden-W\"urttemberg-Stiftung for the financial support by the Eliteprogramm for Postdocs.

\clearpage

\section{Supplementary Material: Observation of Rydberg Blockade Induced by a Single Ion}

\subsection{Photo-ionization protocol}

\begin{figure}[!t]
\centering
	\includegraphics[width=\columnwidth]{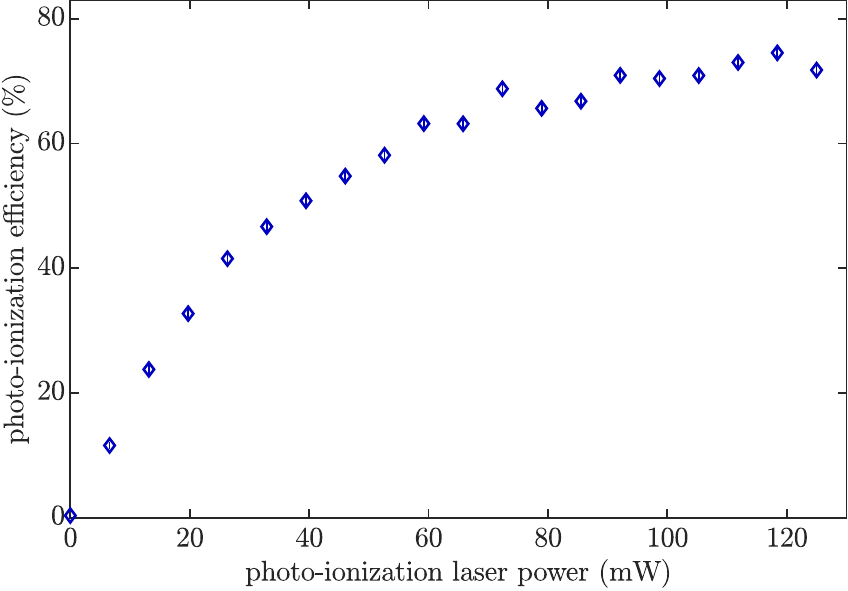}
	\caption{Photo-ionization efficiency of a single Rydberg excitation ($n=51$) as a function of laser power in the 1010.2 nm beam for a pulse length of 200 ns. Error bars denote one standard deviation.}
	\label{FigSup1}
\end{figure}

The photo-ionization laser beams co-propagate with the Rydberg excitation laser at 1015 nm wavelength and are also focused through the high-NA aspheric lens into the atomic ensemble. The first branch of the V-type photo-ionization scheme, which couples the $|nS_{1/2},m_J=+1/2 \rangle$ Rydberg state resonantly to the $6P_{3/2}$ level, is circularly polarized to drive $\sigma^-$ transitions. The second branch at 1010.2 nm wavelength has the same circular polarization. The limiting factor for fast ionization is the photo-ionization rate of the $6P_{3/2}$ state, which depends on the photo-ionization cross section and the intensity of the ionization laser \cite{Courtade2004MAT}. For our strongly focused laser beams, we achieve up to 70~\% photo-ionization efficiency within a pulse of 200 ns. The measured ionization probability as a function of laser power in the 1010.2 nm beam is shown in Fig.~\ref{FigSup1}.

\subsection{Ion and Rydberg atom detection}

As described in the main article, a two-step electric field pulse is applied in the experiment for individual detection of the produced ion and the Rydberg atom on the MCP. Here, we provide details on our method to extract the post-selected Rydberg signal $N_{\rm{R}}$ for which ion creation and detection was successful.

\begin{figure}[!t]
\centering
	\includegraphics[width=\columnwidth]{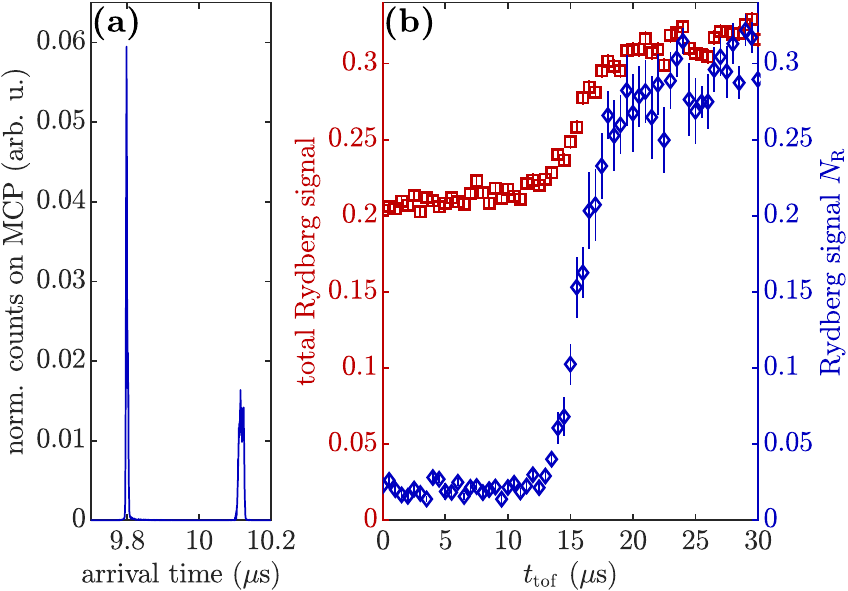}
	\caption{(a) Arrival time distribution of the counts on the MCP for the data set shown in Fig.~1(c) of the main article, taking into account realizations with $t_{\rm{tof}} \leq 10 \mu \rm{s}$. The field-ionized Rydberg atoms arrive at $9.8 \, \mu \rm{s}$, whereas the ions arrive at $10.1 \, \mu \rm{s}$. (b) Comparison of the post-selected Rydberg signal $N_{\rm{R}}$ (diamonds), shown in Fig.~1(c) of the main article, with the total Rydberg signal (squares), which also includes events where no ion was created or detected.}
	\label{FigSup2}
\end{figure} 

Fig.~\ref{FigSup2}(a) shows the arrival time distribution of detected counts on the MCP at the example of the data set shown in Fig.~1(c) of the main article. For the chosen parameters of the two-step electric field pulse, i.e. step size and step duration, the field-ionized Rydberg atoms arrive after $9.8 \, \mu \rm{s}$ at the MCP, while the arrival time of the ions is around $10.1 \, \mu \rm{s}$. The fact that the field-ionized Rydberg atoms pass the ions on their way to the detector and eventually arrive first is a consequence of our electrode geometry in combination with the field pulse sequence. Also note that the first peak includes the signal stemming from both Rydberg excitation pulses, i.e. the first pulse prior to ion creation and the second pulse for probing the blockade. The clear temporal separation of the ion and Rydberg atom signal allows for post-selecting events for which the ion was successfully created and detected to evaluate $N_{\rm{R}}$.

The implication of extracting the conditional Rydberg signal $N_{\rm{R}}$ is illustrated in Fig.~\ref{FigSup2}(b). Here, the data set for $N_{\rm{R}}$ as a function of $t_{\rm{tof}}$ shown in Fig.~1(c) of the main article is compared to the non-conditional Rydberg signal, i.e. the total Rydberg signal obtained without applying the post-selection procedure. Clearly, the contrast for the latter is significantly reduced, which can be understood from the combined effects of finite photo-ionization and MCP detection efficiency.

\subsection{Numerical calculations of the Rydberg excitation dynamics and definition of the excitation bandwidth}

In the main article, we extract $R_{\rm{b}}$ from $E^{\star}$ obtained from the error-function fits to the data in Fig.~2 using simple kinetics for the accelerated motion of the ion,
\begin{equation}
R_{\rm{b}} = e E^{\star} t_{\rm{tof}}^2 / (2m) \, .
\label{EqSup1}
\end{equation}
Here, we verify this approach by numerical calculations of the Rydberg excitation dynamics in the presence of the ion. Specifically, these calculations quantify that the extracted center position of the error-function $E^{\star}$ and the blockade radius $R_{\rm{b}}$ are indeed connected via Eq.~\ref{EqSup1}. In order to show this, we numerically solve the time evolution for a single two-level-atom with a time-dependent energy shift of the excited state $V(R)$ according to the accelerated motion of the ion in an electric field $E$. In the simulation, the Rydberg excitation pulse length is set to reflect the experimentally measured Fourier-limited bandwidth $\Gamma$, which denotes the half width at half maximum of the Rydberg excitation spectrum in the absence of the ion.

\begin{figure}[!t]
\centering
	\includegraphics[width=\columnwidth]{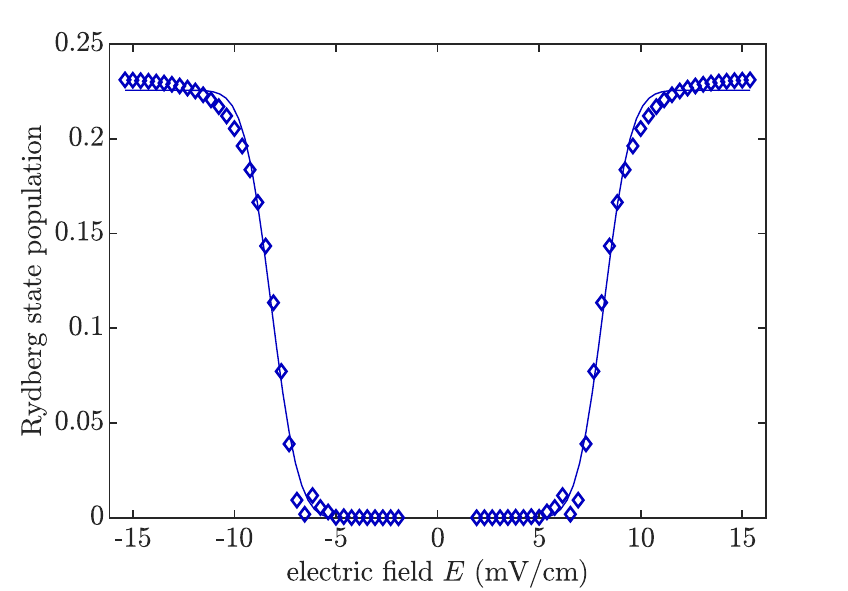}
	\caption{Numerical simulation of the Rydberg excitation dynamics in the presence of the ion. The simulated Rydberg state population (symbols) after the excitation pulse is shown as a function of electric field $E$ for $n=90$. The Rabi frequency is set to $2 \pi \times 400 \, \rm{kHz}$ to reflect the mean Rydberg population in the experiment. The solid line is an error-function fit to the simulation results to extract $R_{\rm{b}}$.}
	\label{FigSup3}
\end{figure}

An exemplary result of the simulated population in the excited state at the end of the Rydberg excitation pulse is shown in Fig.~\ref{FigSup3} as a function of $E$ for the case of $n=90$. The numerical simulation well resembles the shape of the observed experimental signal in Fig.~2. As in the experiment, we fit the numerical data with an error-function to extract $E^{\star}$. We find that $R_{\rm{b}}$ obtained from the fit result for $E^{\star}$ using Eq.~\ref{EqSup1} agrees very well with the theoretical prediction given by Eq.~2 in the main article. We have checked that this holds for the entire range of principal quantum numbers investigated in this work.

\subsection{Correction for the blockade radius due to the applied electric field}

We compare the data in Fig.~3 of the main article to an extended theoretical description of the blockade radius $R_{\rm{b}}$ which takes the applied electric field $E_x$ during the measurement into account. As mentioned in the main text, the field causes a systematic shift of $R_{\rm{b}}$ to smaller values, specifically pronounced for large $n$. This is due to the reduced total field strength, i.e. the sum of the Coulomb field of the ion $E_{\rm{ion}}=e/(4 \pi \epsilon_0 R^2)$ and $E_x$, at the position of the Rydberg atom. For a given $E_x$, the pair potential $V(R)$ is modified accordingly and reads
\begin{equation}
V_{\rm{mod}}(R) = - \alpha_{nS} \frac{(E_{\rm{ion}} - |E_x|)^2}{2} \, .
\label{EqSup2}
\end{equation}
This modified potential is compared to the bare $V(R)$ in the inset to Fig.~3 in the main article. The blockade radius for $V_{\rm{mod}}(R)$ is obtained by equating the pair interaction with the excitation bandwidth $\Gamma$ and takes the form
\begin{equation}
R_{\rm{b}}^{\rm{mod}} = \left( \frac{e  \alpha_{nS}}{4 \pi \epsilon_0 (|E_x|  \alpha_{nS} + \sqrt{2  \alpha_{nS} \Gamma})} \right)^{1/2} \, .
\label{EqSup2}
\end{equation}
In the experiment, the modified blockade radius is measured for $E_x = E^{\star}$. The measured $E^{\star}$, however, depends itself on $R_{\rm{b}}^{\rm{mod}}$. Consequently, the theoretical prediction for $R_{\rm{b}}^{\rm{mod}}$ and $E^{\star}$ needs to be calculated in an iterative and self-consistent way. For a given principal quantum number $n$, one may start from the bare value of $R_{\rm{b}}$ given by Eq.~2 in the main text. The corresponding $E^{\star}$ obtained from the ion's trajectory, Eq.~\ref{EqSup1}, can then be plugged into Eq.~\ref{EqSup2}, which delivers a corrected value for $R_{\rm{b}}$. After a few iterations, this procedure yields self-consistent values for the corrected $R_{\rm{b}}$ and $E^{\star}$, which we compare in Fig.~3 of the main article to the experimental data. Note that we have verified this approach with the numerical results of the excitation dynamics described in the previous paragraph by adding the field $E_x$ into the simulations.

\clearpage

\end{document}